\documentclass{PoS}
\usepackage{amsmath}
\usepackage{xspace}

\newcommand{\ATLAS}{ATLAS\xspace}
\newcommand{\CMS}{CMS\xspace}
\newcommand{\ALEPH}{ALEPH\xspace}

\newcommand{\MCatNLO}{M\protect\scalebox{0.8}{C}@N\protect\scalebox{0.8}{LO}\xspace}

\newcommand{\NLOPS}{N\protect\scalebox{0.8}{LO}P\protect\scalebox{0.8}{S}\xspace}
\newcommand{\MENLOPS}{ME\protect\scalebox{0.8}{NLO}PS\xspace}
\newcommand{\MEPSatNLO}{M\protect\scalebox{0.8}{E}P\protect\scalebox{0.8}{S}@N\protect\scalebox{0.8}{LO}\xspace}

\newcommand{\Sherpa}{S\protect\scalebox{0.8}{HERPA}\xspace}
\newcommand{\Comix}{C\protect\scalebox{0.8}{OMIX}\xspace}

\newcommand{\Amegic}{A\protect\scalebox{0.8}{MEGIC++}\xspace}
\newcommand{\CSS}{C\protect\scalebox{0.8}{SS}\xspace}

\newcommand{\BlackHat}{B\protect\scalebox{0.8}{LACK}H\protect\scalebox{0.8}{AT}\xspace}
\newcommand{\OpenLoops}{O\protect\scalebox{0.8}{PEN}L\protect\scalebox{0.8}{OOPS}\xspace}

\newcommand{\MET}{\slash\!\!\!\!\! E_\perp}

\pdfoutput=1

\title{\vspace{-3.5cm}{\small IPPP/13/94, DCPT/13/188, MCNET--13--20, LPN13--093, SLAC--PUB--15813}\\\vspace{2cm}
       Merging of matrix elements and parton showers at NLO accuracy}

\ShortTitle{Merging of matrix elements and parton showers at NLO accuracy}

\author{\speaker{Marek Sch\"onherr}\\
        Institute for Particle Physics Phenomenology,
        Durham University, Durham DH1 3LE, UK}

\author{Stefan H\"oche\\
        SLAC National Accelerator Laboratory, 
        Menlo Park, CA 94025, USA}

\author{Frank Krauss\\
        Institute for Particle Physics Phenomenology,
        Durham University, Durham DH1 3LE, UK}

\author{Frank Siegert\\
        Institut f\"ur Kern- und Teilchenphysik,
        TU Dresden, D--01062 Dresden, Germany}

\abstract{The merging of matrix elements and parton showers is an 
          established calculational tool for the description of 
          multi-jet final states at hadron colliders. These methods 
          have recently been promoted to next-to-leading order 
          accuracy in the description of hard well separated jets. 
          This talk introduces such a method and discusses its 
          application to phenomenologically relevant signal and 
          background processes. The systematic assessment of its 
          theoretical uncertainty is a prime focus.}

\FullConference{The European Physical Society Conference on High Energy Physics -EPS-HEP2013\\
		18-24 July 2013\\
		Stockholm, Sweden}

\begin{document}

\section{Introduction}

Many experimentally important observables at recent and present 
particle colliders are dominated by final states of (multiple) large 
multiplicities of not necessarily well separated leptons, photons, 
hadrons and/or jets. Thus, achieving an accurate theoretical description 
of such observables necessitates not only matching higher terms in an 
expansion in terms of $\alpha_s$, such as a next-to-leading order 
calculation, improving the description of short-distance physics, to a 
parton shower resummation, accurately describing parton evolution at 
small scales, for a process of fixed parton multiplicity, but also 
merging such matched calculations for subsequent multiplicities without 
the loss of their respective accuracies.

In the CKKW \cite{Catani:2001cc} type of formalisms a few approaches have 
been formulated recently \cite{Lavesson:2008ah,Hoeche:2012yf,
  Gehrmann:2012yg,Platzer:2012bs,Lonnblad:2012ix} differing in their 
choice of tools and treatment of the overlap of the individual input 
calculations. The following reviews recent results with the implementation 
of this so-called \MEPSatNLO method in the event generator \Sherpa.

\section{Recent results}

In this section a few results obtained recently with the implementation 
of the \MEPSatNLO method in the \Sherpa \cite{Gleisberg:2008ta} event 
generator are exhibited. They use \Sherpa's tree-level matrix element 
generators \Amegic \cite{Krauss:2001iv,Gleisberg:2003ue} and \Comix 
\cite{Gleisberg:2008fv}, its Catani-Seymour dipole subtraction 
implementation \cite{Catani:1996vz,Gleisberg:2007md} and \Sherpa's 
CS-dipole shower, \CSS, \cite{Schumann:2007mg}. For all processes shown 
here, one-loop matrix elements were provided by \BlackHat 
\cite{Berger:2008sj} or \OpenLoops \cite{Cascioli:2011va}. \Sherpa further 
comprises modules to compute the effects of multi-parton interactions 
\cite{Alekhin:2005dx}, hadronisation corrections \cite{Winter:2003tt}, 
hadron decays and higher-order QED corrections \cite{Schonherr:2008av}.

The shown \MEPSatNLO calculations are constructed from \NLOPS calculations 
for the individual parton multiplicities, matched using the methods of 
\cite{Hoeche:2011fd}\footnote{The applicability of this method to 
processes of most general colour structures was demonstrated in 
\cite{Hoeche:2012ft,Hoeche:2012fm,Cascioli:2013era}.}, according to 
the algorithm developed in \cite{Hoeche:2012yf,Gehrmann:2012yg,
  Hoeche:2013mua,Cascioli:2013gfa}. The \MENLOPS technique -- the presented 
results use an implementation according to \cite{Gehrmann:2012yg,Hoche:2010pf,
  Hoeche:2010kg} -- is a special case of a \MEPSatNLO calculation where 
only the lowest multiplicity is calculated at \NLOPS accuracy and all 
higher multiplicities are merged at leading order accuracy.

\begin{figure}[p]
  \begin{minipage}{0.47\textwidth}
    \lineskip-1.6pt
    \includegraphics[width=\textwidth]{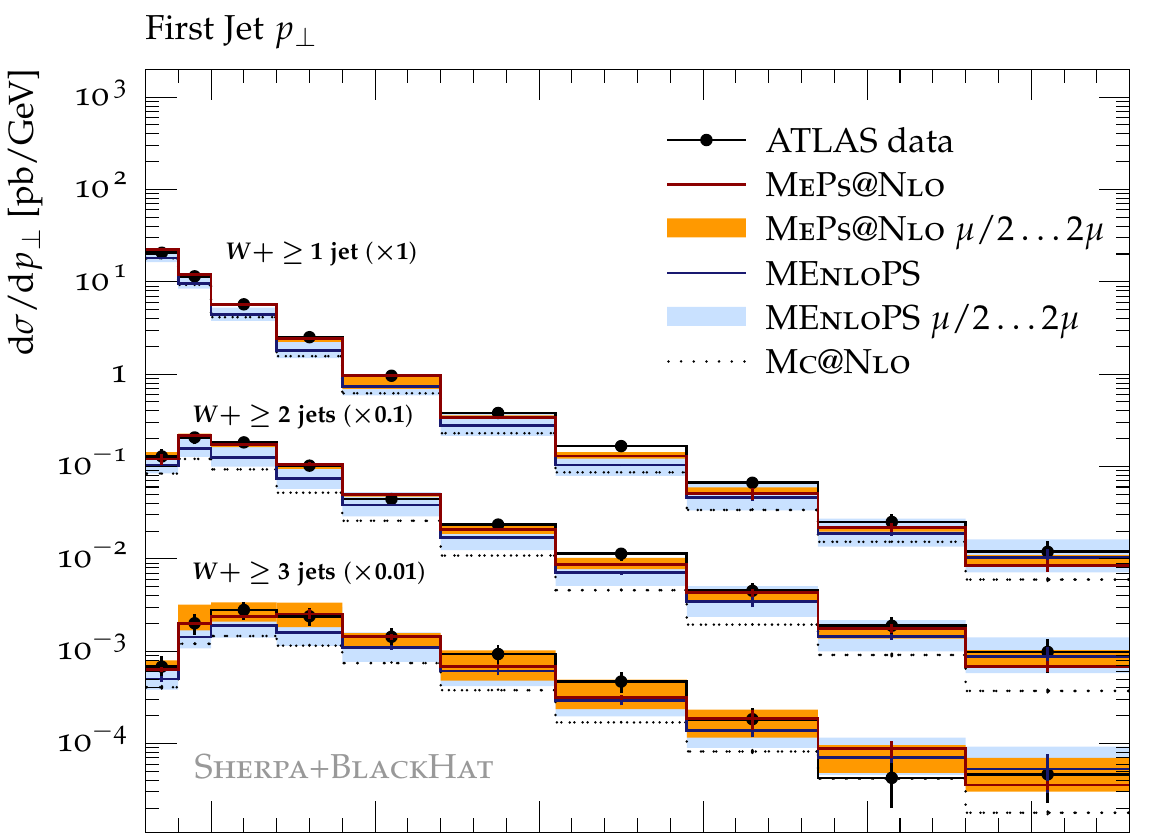}
    \includegraphics[width=\textwidth]{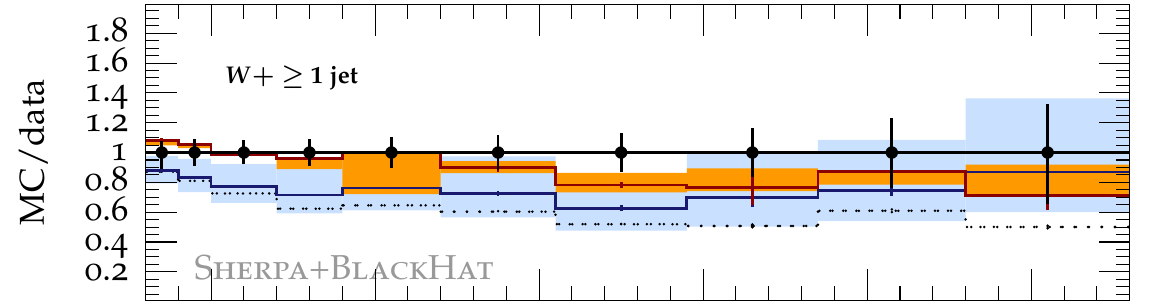}
    \includegraphics[width=\textwidth]{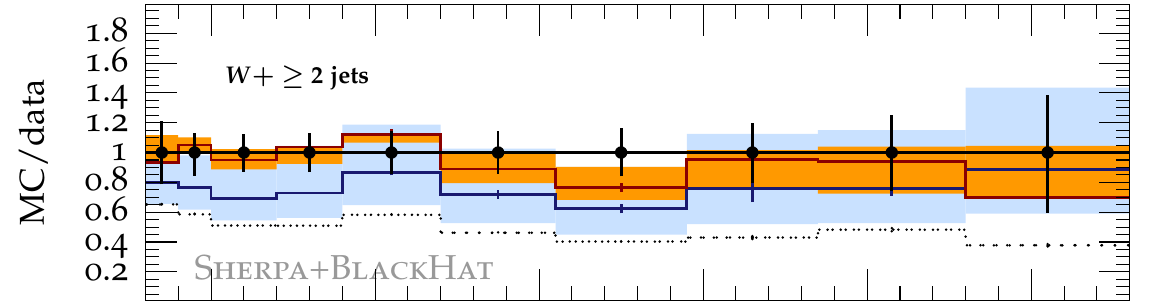}
    \includegraphics[width=\textwidth]{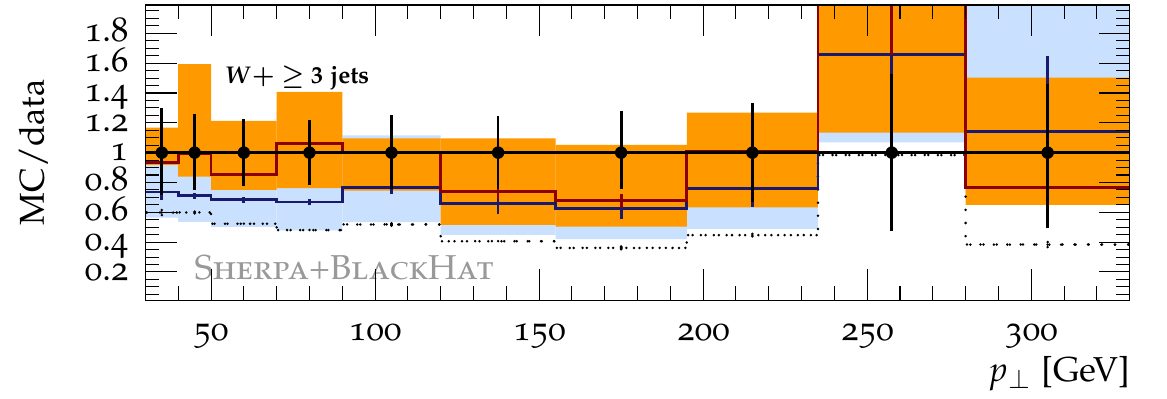}
  \end{minipage}\hfill
  \begin{minipage}{0.47\textwidth}
    \lineskip-1.6pt
    \includegraphics[width=\textwidth]{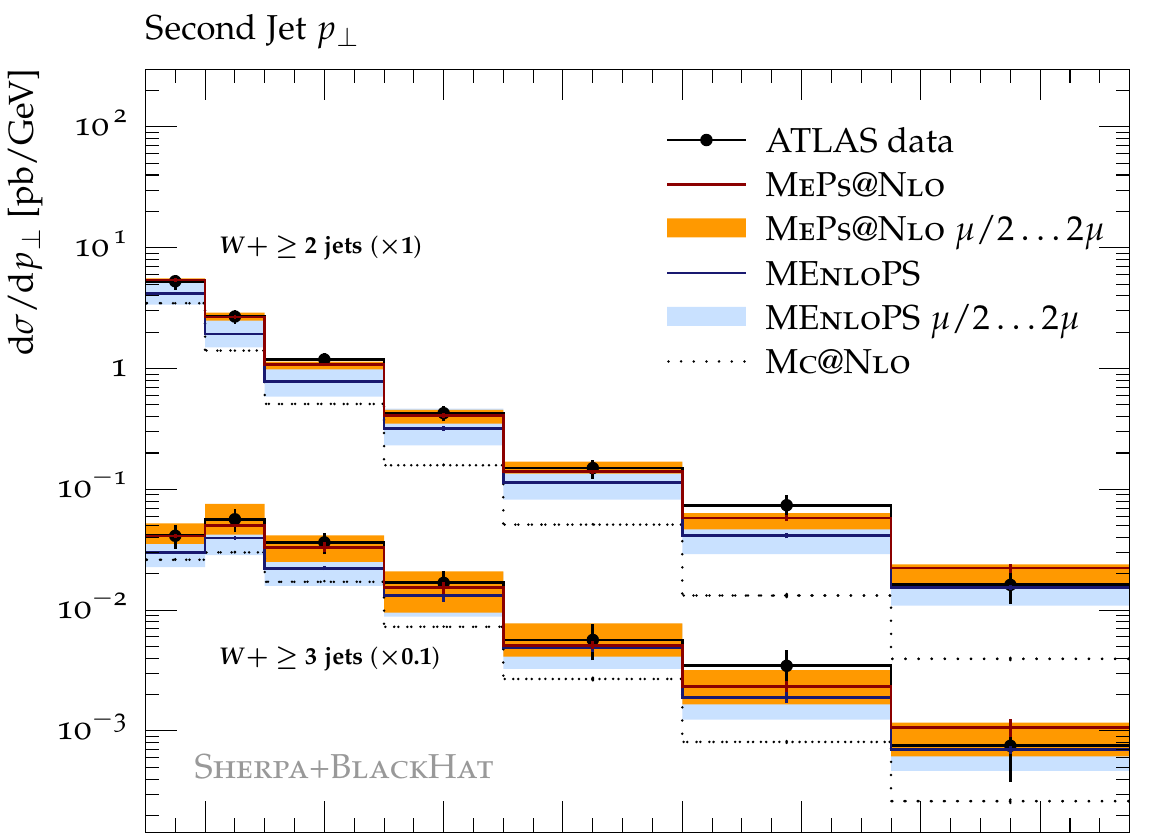}
    \includegraphics[width=\textwidth]{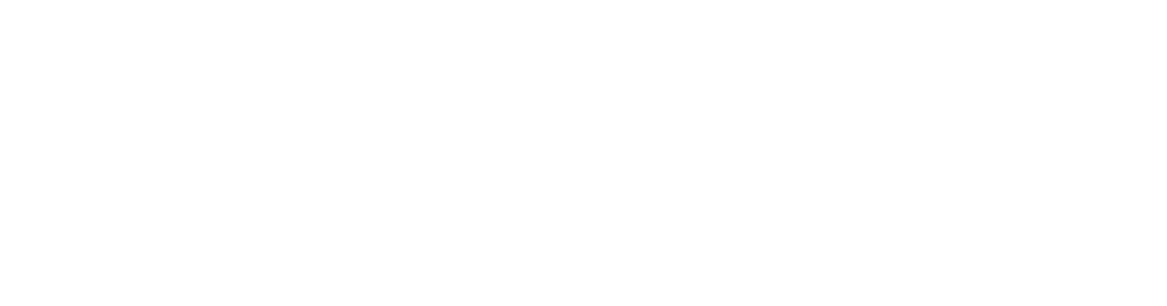}
    \includegraphics[width=\textwidth]{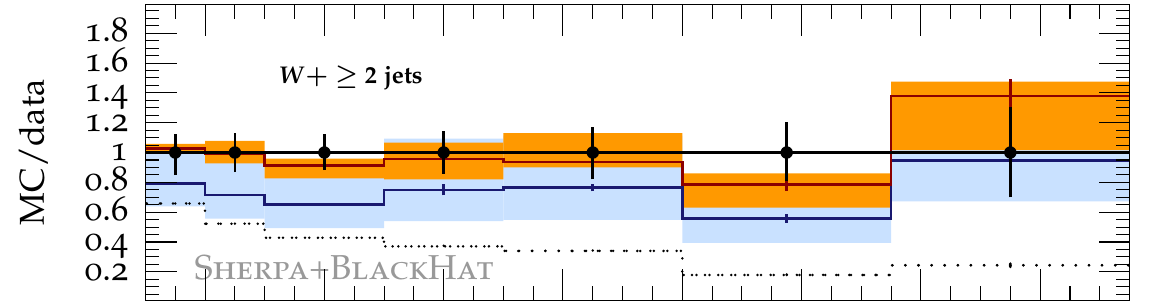}
    \includegraphics[width=\textwidth]{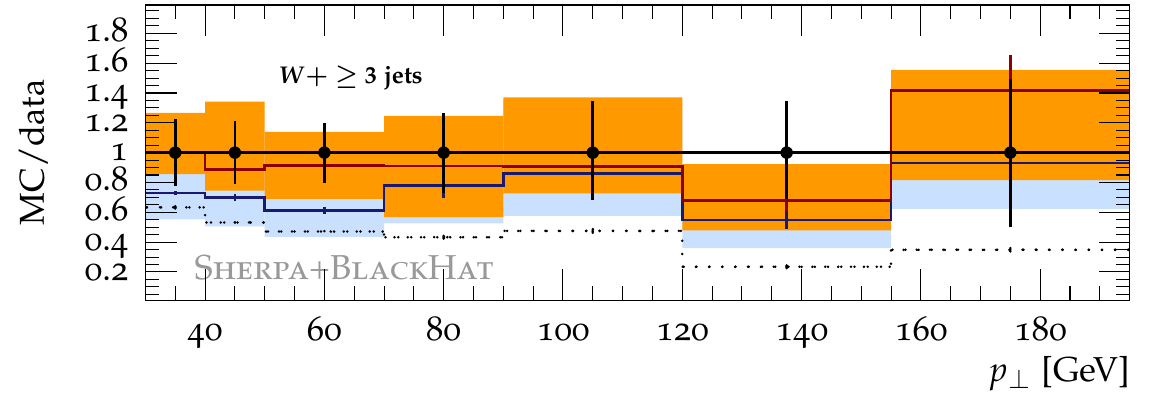}
  \end{minipage}
  \caption
  {
    Transverse momentum of the leading (left) and subleading (right) 
    jet in $pp\to W+\text{jets}$ events with at least one, two or 
    three jets compared to \ATLAS data \cite{Aad:2012en}. Displayed 
    are the results of a \MEPSatNLO (red), \MENLOPS (blue) and \MCatNLO 
    (black dotted) simulation of the inclusive process. For both 
    multijet merged predictions the renormalisation and factorisation 
    uncertainties are shown.
    \label{fig:w}
  }
\end{figure}

\begin{figure}[p]
  \begin{center}
    \includegraphics[width=0.47\textwidth]{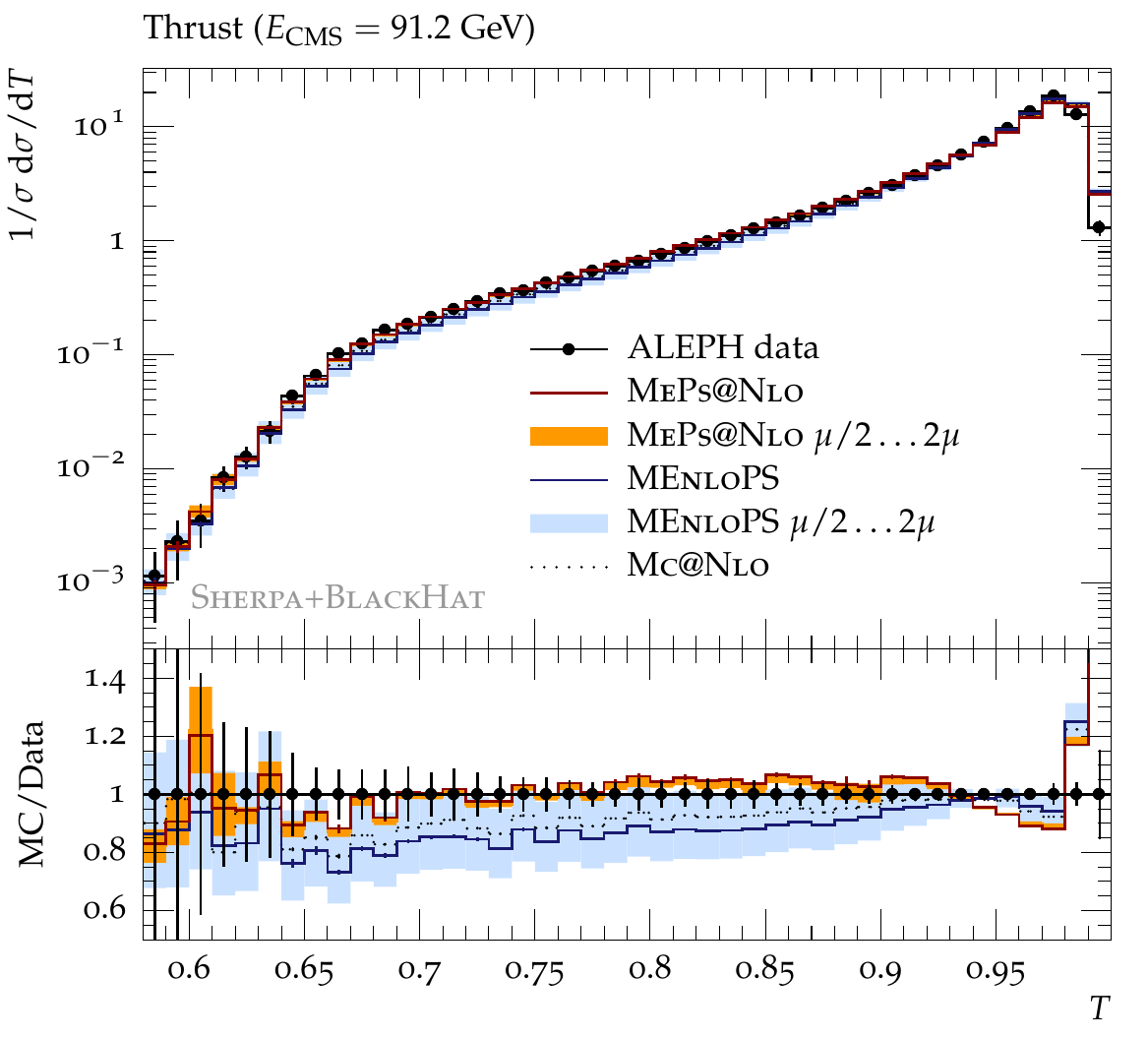}\hfill
    \includegraphics[width=0.47\textwidth]{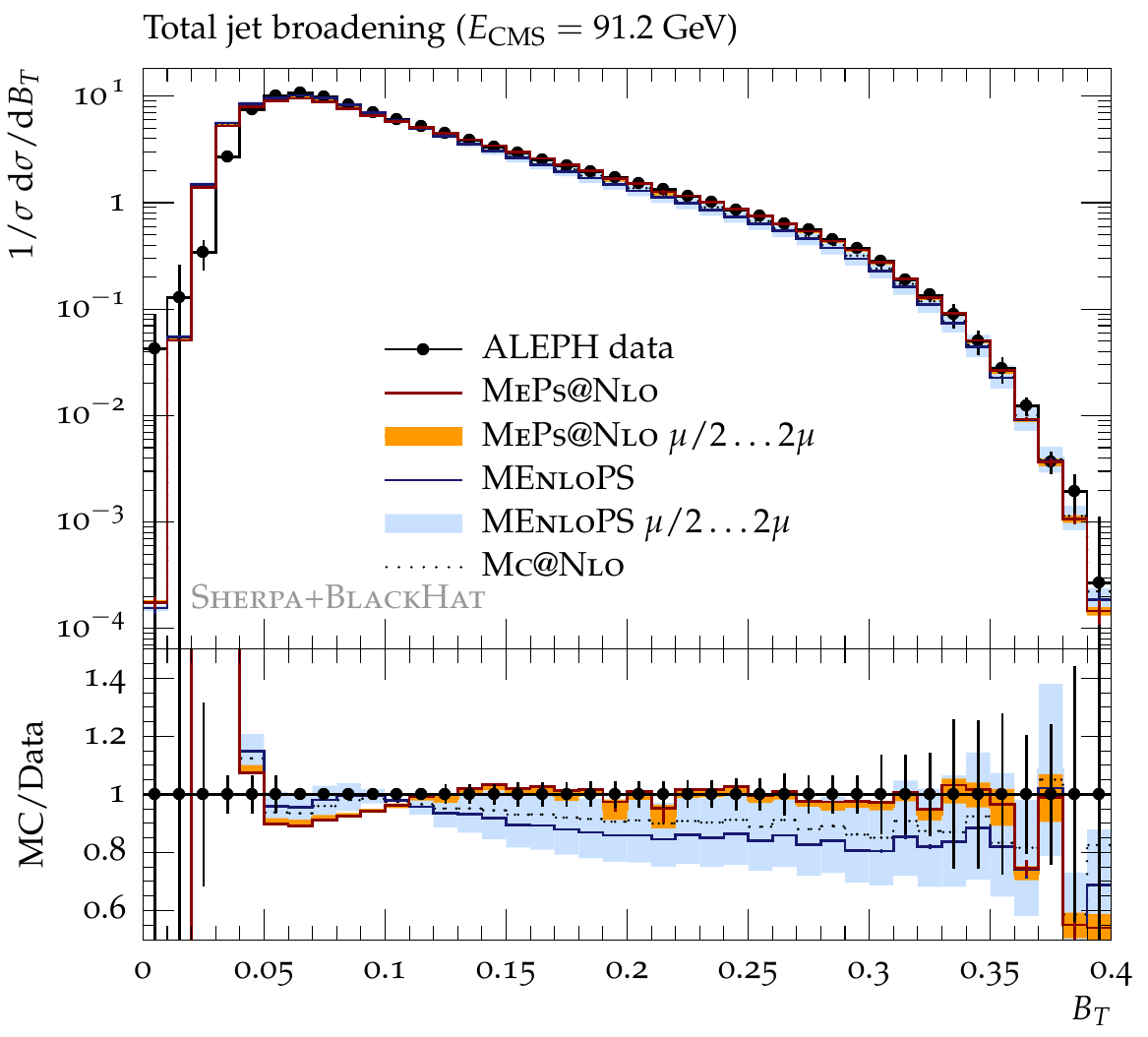}
  \end{center}
  \caption
  {
    Thrust (left) and total jet broadening (right) in $e^+e^-\to\text{jets}$ 
    events compared to \ALEPH data \cite{Heister:2003aj}. Displayed 
    are the results of a \MEPSatNLO (red), \MENLOPS (blue) and \MCatNLO 
    (black dotted) simulation of the inclusive process. For both 
    multijet merged predictions the renormalisation and factorisation 
    uncertainties are shown.
    \label{fig:ee}
  }
\end{figure}

\begin{figure}[p]
  \begin{minipage}{0.47\textwidth}
    \lineskip-1.6pt
    \includegraphics[width=\textwidth]{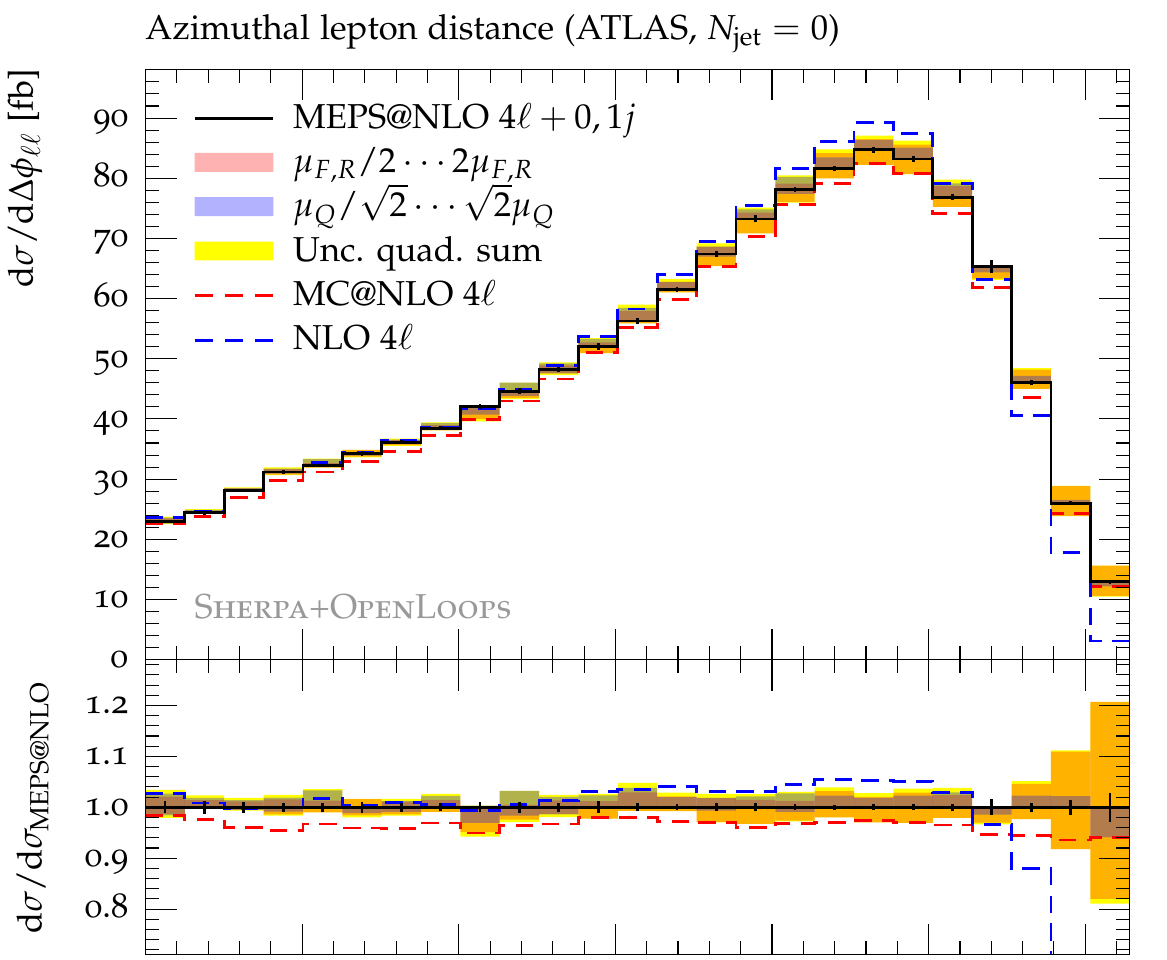}
    \includegraphics[width=\textwidth]{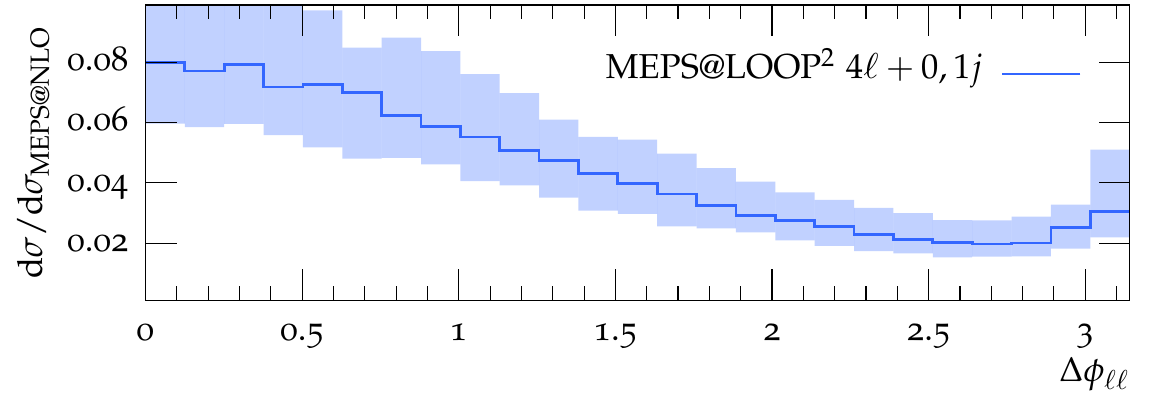}
  \end{minipage}\hfill
  \begin{minipage}{0.47\textwidth}
    \lineskip-1.6pt
    \includegraphics[width=\textwidth]{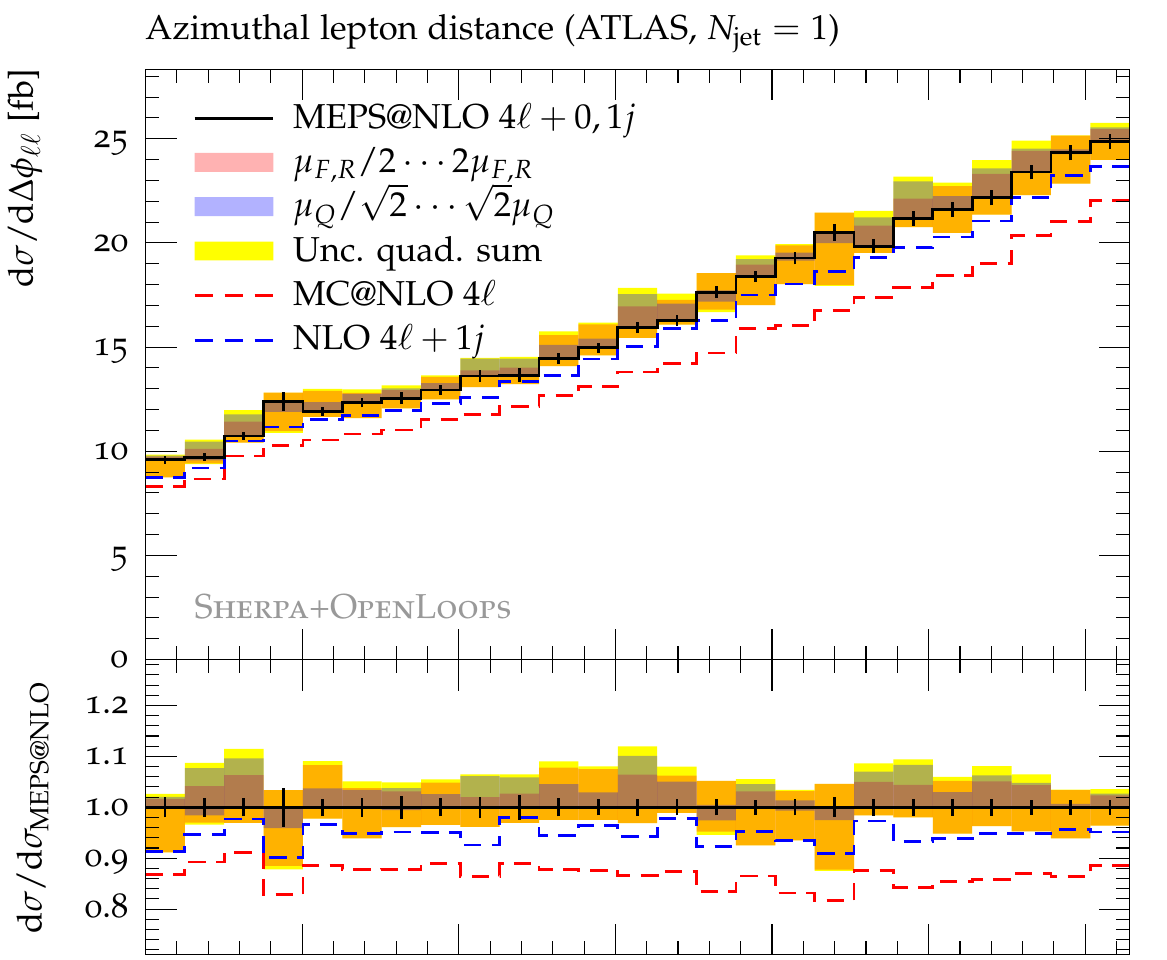}
    \includegraphics[width=\textwidth]{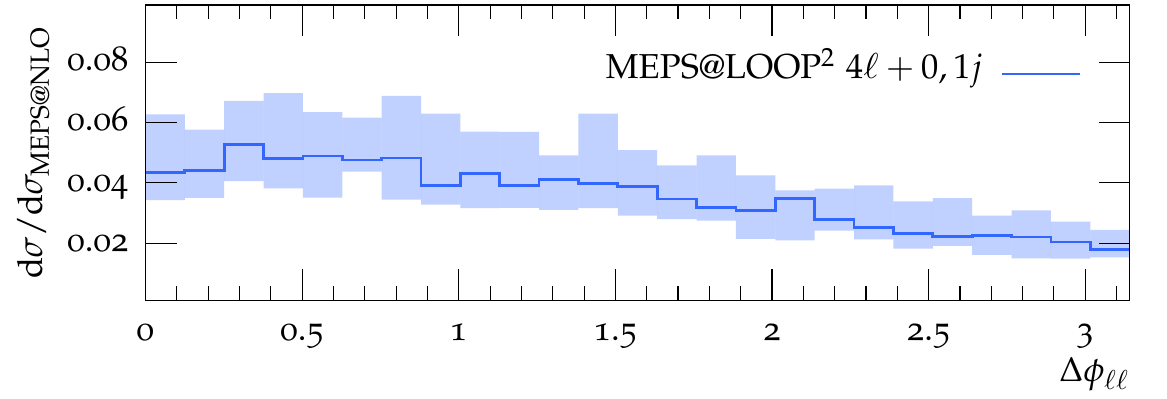}
  \end{minipage}
  \caption
  {
    Azimuthal separation of the lepton pair in $pp\to\ell^+\ell^-\!+\MET+X$ 
    in the exclusive zero (left) and one (right) jet selection under 
    typical cuts used in Higgs searches. Displayed are the \MEPSatNLO 
    (black) prediction, with its associated renormalisation and 
    factorisation scale uncertainties as well as the resummation scale 
    uncertainty, the inclusive \MCatNLO prediction and the fixed-order 
    NLO prediction. The lower panel shows the contribution of loop 
    induced processes to the total result.
    \label{fig:4l_dphi}
  }
\end{figure}

\begin{figure}[p]
  \begin{minipage}{0.47\textwidth}
    \lineskip-1.6pt
    \includegraphics[width=\textwidth]{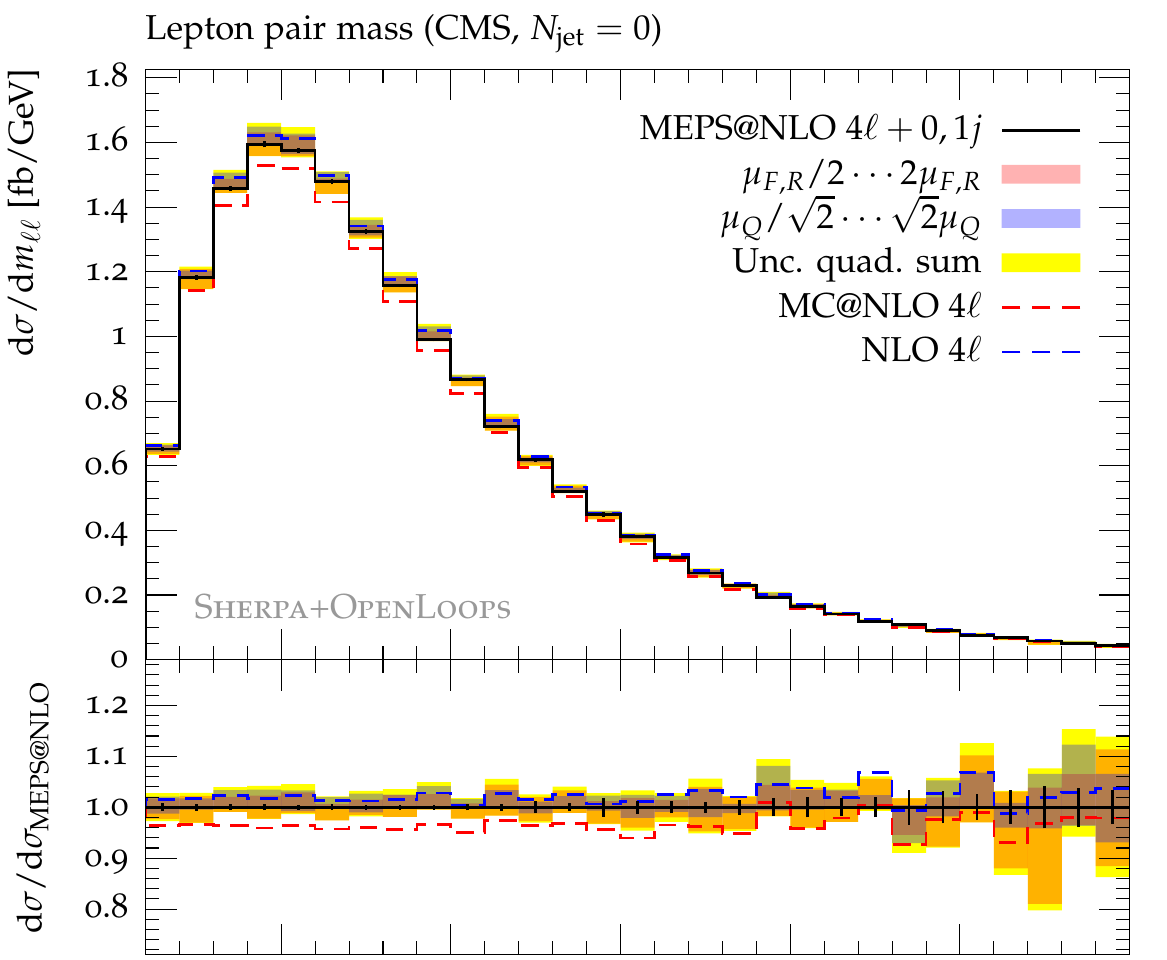}
    \includegraphics[width=\textwidth]{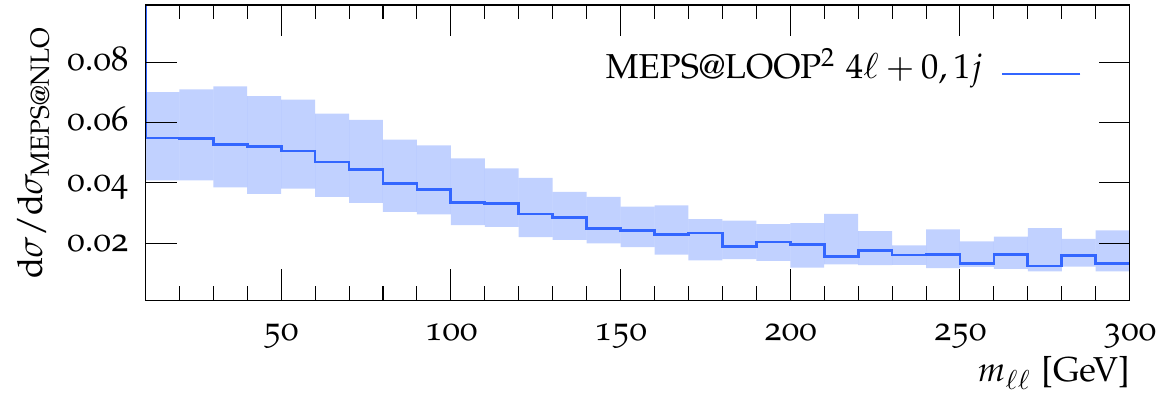}
  \end{minipage}\hfill
  \begin{minipage}{0.47\textwidth}
    \lineskip-1.6pt
    \includegraphics[width=\textwidth]{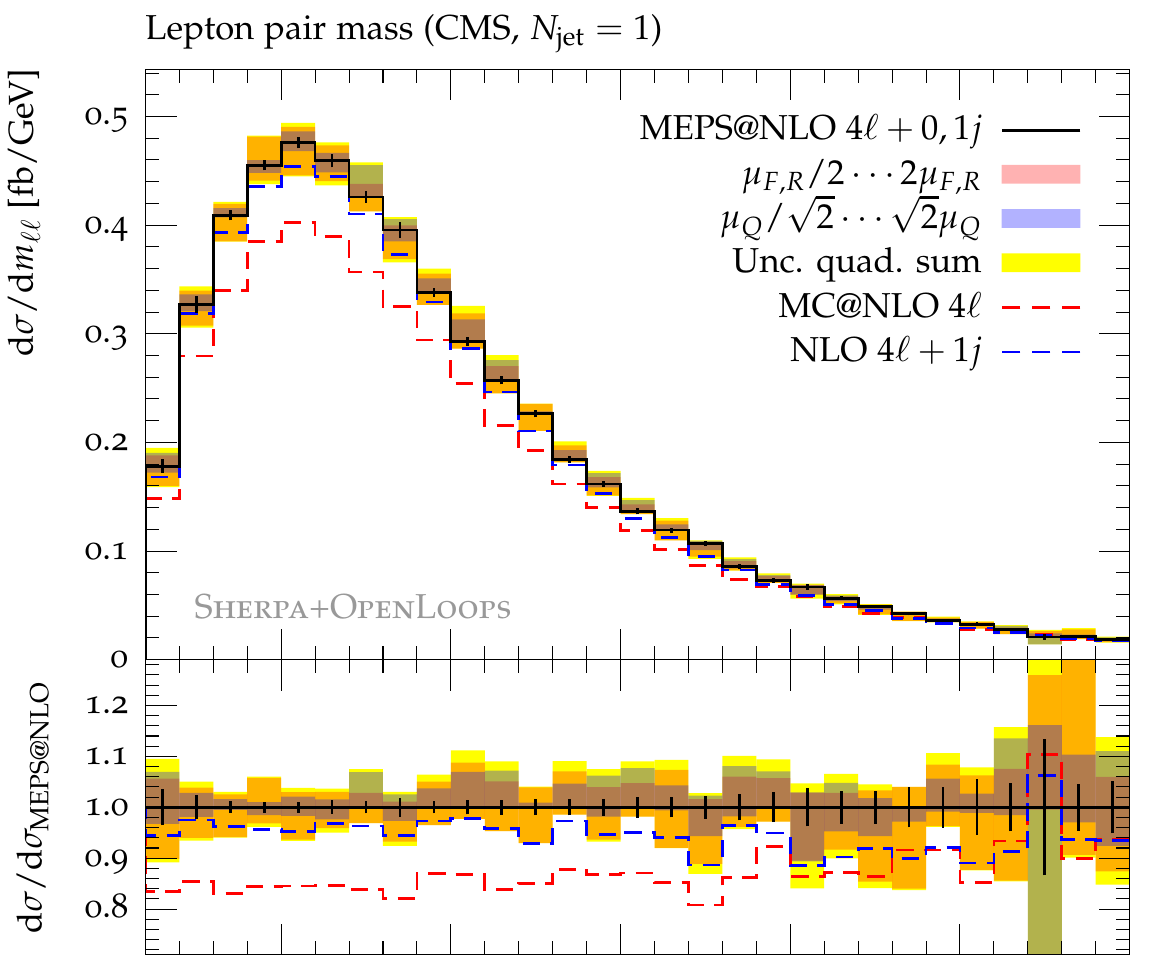}
    \includegraphics[width=\textwidth]{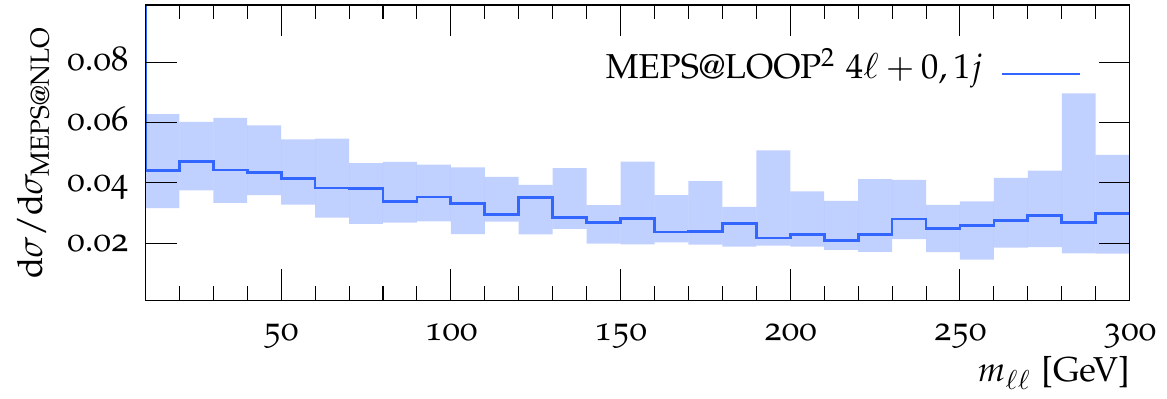}
  \end{minipage}
  \caption
  {
    Invariant mass of the lepton pair in $pp\to\ell^+\ell^-\!+\MET+X$ 
    in the exclusive zero (left) and one (right) jet selection under 
    typical cuts used in Higgs searches. Displayed are the \MEPSatNLO 
    (black) prediction, with its associated renormalisation and 
    factorisation scale uncertainties as well as the resummation scale 
    uncertainty, the inclusive \MCatNLO prediction and the fixed-order 
    NLO prediction. The lower panel shows the contribution of loop 
    induced processes to the total result.
    \label{fig:4l_mass}
  }
\end{figure}

Fig.\ \ref{fig:w} displays results obtained with the afore mentioned 
calculations for $pp\to W+\text{jets}$ production \cite{Hoeche:2012yf}. 
The $pp\to W$, $pp\to Wj$ and $pp\to Wjj$ contributions were calculated 
at \NLOPS accuracy. Additionally, the processes $pp\to Wjjj$ and 
$pp\to Wjjjj$ were merged on top of that at leading-order accuracy.
The results are compared to experimental data from the \ATLAS collaboration 
\cite{Aad:2012en}. Good agreement is found and, of equal importance, 
the theoretical uncertainty is reduced using next-to-leading order 
merging. Similar observations are made when comparing a \MEPSatNLO 
simulation for $e^+e^-\to\text{jets}$ \cite{Gehrmann:2012yg} to data 
taken by the \ALEPH collaboration \cite{Heister:2003aj}, as displayed 
in Fig.\ \ref{fig:ee}. This calculation computes $e^+e^-\to jj$, 
$e^+e^-\to jjj$ and $e^+e^-\to jjjj$ at next-to-leading order accuracy, 
again merging two more jets on top of that at leading order accuracy.

Finally, Figs.\ \ref{fig:4l_dphi} and \ref{fig:4l_mass} show results 
for \MEPSatNLO calculations for $pp\to\ell^+\ell^-\!+\MET+X$ production
\cite{Cascioli:2013gfa}, merging $pp\to\ell^+\ell^-\nu\nu$ and 
$pp\to\ell^+\ell^-\nu\nu j$ at \NLOPS accuracy. Included are also the 
loop-induced processes $gg\to\ell^+\ell^-\nu\nu$, 
$gg\to\ell^+\ell^-\nu\nu g$ and $gq\to\ell^+\ell^-\nu\nu q$. 
Emphasis here is put of course onto observables relevant for Higgs 
boson measurements as used by \ATLAS and \CMS, as usual divided into 
bins of exclusive jet multiplicity. The distributions show that both an 
improved central value and a small theoretical uncertainty, as compared 
to pure fixed-order NLO or an inclusive \MCatNLO (generated according 
to \cite{Hoeche:2011fd}), can be achieved.

\section{Conclusions}

The method for merging multiple matrix elements of successive jet 
multiplicities at next-to-leading order accuracy implemented in the 
\Sherpa event generator is a versatile tool that has been shown to 
be applicable to a multitude of processes. It describes the respective 
jet multiplicities at next-to-leading order accuracy which is reflected 
not only in a improved central value but also in smaller theoretical 
uncertainties. At the same time it also preserves the resummation of 
emission scale hierarchies provided by the parton shower. This allows 
to use the \Sherpa Monte Carlo event generator to calculate inclusive 
multijet observables and study their uncertainty due to the truncation 
of the perturbative series in a systematic way.

\bibliographystyle{bib/amsunsrt_mod}
\bibliography{bib/journal}

\begin{thebibliography}{10}

\bibitem{Catani:2001cc}
S.~Catani, F.~Krauss, R.~Kuhn and B.~R. Webber, \emph{{QCD matrix elements +
  parton showers}}, JHEP \textbf{11} (2001),
  \href{http://www.slac.stanford.edu/spires/find/hep/www?eprint=hep-ph/0109231}{063},
   [\href{http://arXiv.org/pdf/hep-ph/0109231}{{\tt hep-ph/0109231}}]. \relax
 \relax
\bibitem{Lavesson:2008ah}
N.~Lavesson and L.~L{\"o}nnblad, \emph{{Extending CKKW-merging to one-loop
  matrix elements}}, JHEP \textbf{12} (2008),
  \href{http://www.slac.stanford.edu/spires/find/hep/www?eprint=0811.2912}{070},
   [\href{http://arXiv.org/pdf/0811.2912}{{\tt arXiv:0811.2912}} [hep-ph]].
  \relax
 \relax
\bibitem{Hoeche:2012yf}
S.~H{\"o}che, F.~Krauss, M.~Sch{\"o}nherr and F.~Siegert, \emph{{QCD matrix
  elements + parton showers: The NLO case}}, JHEP \textbf{1304} (2013),
  \href{http://inspirehep.net/record/1123387}{027},
  [\href{http://arXiv.org/pdf/1207.5030}{{\tt arXiv:1207.5030}} [hep-ph]].
  \relax
 \relax
\bibitem{Gehrmann:2012yg}
T.~Gehrmann, S.~H{\"o}che, F.~Krauss, M.~Sch{\"o}nherr and F.~Siegert,
  \emph{{NLO QCD matrix elements + parton showers in $e^+e^-\to$hadrons}}, JHEP
  \textbf{1301} (2013),
  \href{http://www.slac.stanford.edu/spires/find/hep/www?eprint=1207.5031}{144},
   [\href{http://arXiv.org/pdf/1207.5031}{{\tt arXiv:1207.5031}} [hep-ph]].
  \relax
 \relax
\bibitem{Platzer:2012bs}
S.~Pl\"atzer, \emph{{Controlling inclusive cross sections in parton shower +
  matrix element merging}}, JHEP \textbf{1308} (2013),
  \href{http://www.slac.stanford.edu/spires/find/hep/www?eprint=1211.5467}{114},
   [\href{http://arXiv.org/pdf/1211.5467}{{\tt arXiv:1211.5467}} [hep-ph]].
  \relax
 \relax
\bibitem{Lonnblad:2012ix}
L.~L{\"o}nnblad and S.~Prestel, \emph{{Merging Multi-leg NLO Matrix Elements
  with Parton Showers}}, JHEP \textbf{1303} (2013),
  \href{http://inspirehep.net/record/1205021}{166},
  [\href{http://arXiv.org/pdf/1211.7278}{{\tt arXiv:1211.7278}} [hep-ph]].
  \relax
 \relax
\bibitem{Gleisberg:2008ta}
T.~Gleisberg, S.~H{\"o}che, F.~Krauss, M.~Sch\"{o}nherr, S.~Schumann,
  F.~Siegert and J.~Winter, \emph{{Event generation with \Sherpa 1.1}}, JHEP
  \textbf{02} (2009), \href{http://inspirebeta.net/record/803708}{007},
  [\href{http://arXiv.org/pdf/0811.4622}{{\tt arXiv:0811.4622}} [hep-ph]].
  \relax
 \relax
\bibitem{Krauss:2001iv}
F.~Krauss, R.~Kuhn and G.~Soff, \emph{{AMEGIC++ 1.0: A Matrix Element Generator
  In C++}}, JHEP \textbf{02} (2002),
  \href{http://www.slac.stanford.edu/spires/find/hep/www?eprint=hep-ph/0109036}{044},
   [\href{http://arXiv.org/pdf/hep-ph/0109036}{{\tt hep-ph/0109036}}]. \relax
 \relax
\bibitem{Gleisberg:2003ue}
T.~Gleisberg, F.~Krauss, K.~T. Matchev, A.~Sch{\"a}licke, S.~Schumann and
  G.~Soff, \emph{{Helicity formalism for spin-2 particles}}, JHEP \textbf{09}
  (2003),
  \href{http://www.slac.stanford.edu/spires/find/hep/www?eprint=hep-ph/0306182}{001},
   [\href{http://arXiv.org/pdf/hep-ph/0306182}{{\tt hep-ph/0306182}}]. \relax
 \relax
\bibitem{Gleisberg:2008fv}
T.~Gleisberg and S.~H{\"o}che, \emph{{Comix, a new matrix element generator}},
  JHEP \textbf{12} (2008), \href{http://inspirehep.net/record/793879}{039},
  [\href{http://arXiv.org/pdf/0808.3674}{{\tt arXiv:0808.3674}} [hep-ph]].
  \relax
 \relax
\bibitem{Catani:1996vz}
S.~Catani and M.~H. Seymour, \emph{{A general algorithm for calculating jet
  cross sections in NLO QCD}}, Nucl. Phys. \textbf{B485} (1997),
  \href{http://www.slac.stanford.edu/spires/find/hep/www?eprint=hep-ph/9605323}{291--419},
   [\href{http://arXiv.org/pdf/hep-ph/9605323}{{\tt hep-ph/9605323}}]. \relax
 \relax
\bibitem{Gleisberg:2007md}
T.~Gleisberg and F.~Krauss, \emph{{Automating dipole subtraction for QCD NLO
  calculations}}, Eur. Phys. J. \textbf{C53} (2008),
  \href{http://www.slac.stanford.edu/spires/find/hep/www?eprint=arXiv:0709.2881}{501--523},
   [\href{http://arXiv.org/pdf/0709.2881}{{\tt arXiv:0709.2881}} [hep-ph]].
  \relax
 \relax
\bibitem{Schumann:2007mg}
S.~Schumann and F.~Krauss, \emph{{A parton shower algorithm based on
  Catani-Seymour dipole factorisation}}, JHEP \textbf{03} (2008),
  \href{http://www.slac.stanford.edu/spires/find/hep/www?eprint=arXiv:0709.1027}{038},
   [\href{http://arXiv.org/pdf/0709.1027}{{\tt arXiv:0709.1027}} [hep-ph]].
  \relax
 \relax
\bibitem{Berger:2008sj}
C.~F. Berger, Z.~Bern, L.~J. Dixon, F.~Febres-Cordero, D.~Forde, H.~Ita, D.~A.
  Kosower and D.~Ma{\^i}tre, \emph{{Automated implementation of on-shell
  methods for one-loop amplitudes}}, Phys. Rev. \textbf{D78} (2008),
  \href{http://inspirebeta.net/record/782271}{036003},
  [\href{http://arXiv.org/pdf/0803.4180}{{\tt arXiv:0803.4180}} [hep-ph]].
  \relax
 \relax
\bibitem{Cascioli:2011va}
F.~Cascioli, P.~Maierhofer and S.~Pozzorini, \emph{{Scattering Amplitudes with
  Open Loops}}, Eur.Phys.J. \textbf{C72} (2012),
  \href{http://inspirehep.net/record/946998}{1889},
  [\href{http://arXiv.org/pdf/1111.5206}{{\tt arXiv:1111.5206}} [hep-ph]].
  \relax
 \relax
\bibitem{Alekhin:2005dx}
\href{http://www.slac.stanford.edu/spires/find/hep/www?eprint=hep-ph/0601012}{S.~Alekhin
  et~al.}, \emph{{HERA and the LHC - A workshop on the implications of HERA for
  LHC physics: Proceedings Part A}},
  \href{http://arXiv.org/pdf/hep-ph/0601012}{{\tt hep-ph/0601012}}. \relax
 \relax
\bibitem{Winter:2003tt}
J.-C. Winter, F.~Krauss and G.~Soff, \emph{{A modified cluster-hadronisation
  model}}, Eur. Phys. J. \textbf{C36} (2004),
  \href{http://www.slac.stanford.edu/spires/find/hep/www?eprint=hep-ph/0311085}{381--395},
   [\href{http://arXiv.org/pdf/hep-ph/0311085}{{\tt hep-ph/0311085}}]. \relax
 \relax
\bibitem{Schonherr:2008av}
M.~Sch\"{o}nherr and F.~Krauss, \emph{Soft photon radiation in particle decays
  in \Sherpa{}}, JHEP \textbf{12} (2008),
  \href{http://www.slac.stanford.edu/spires/find/hep/www?eprint=arXiv:0810.5071}{018},
   [\href{http://arXiv.org/pdf/0810.5071}{{\tt arXiv:0810.5071}} [hep-ph]].
  \relax
 \relax
\bibitem{Hoeche:2011fd}
S.~H{\"o}che, F.~Krauss, M.~Sch{\"o}nherr and F.~Siegert, \emph{{A critical
  appraisal of NLO+PS matching methods}}, JHEP \textbf{09} (2012),
  \href{http://inspirehep.net/record/944643}{049},
  [\href{http://arXiv.org/pdf/1111.1220}{{\tt arXiv:1111.1220}} [hep-ph]].
  \relax
 \relax
\bibitem{Hoeche:2012ft}
S.~H{\"o}che, F.~Krauss, M.~Sch{\"o}nherr and F.~Siegert, \emph{{W+n-jet
  predictions with MC@NLO in Sherpa}}, Phys.Rev.Lett. \textbf{110} (2013),
  \href{http://inspirehep.net/record/1086175}{052001},
  [\href{http://arXiv.org/pdf/1201.5882}{{\tt arXiv:1201.5882}} [hep-ph]].
  \relax
 \relax
\bibitem{Hoeche:2012fm}
S.~H{\"o}che and M.~Sch{\"o}nherr, \emph{{Uncertainties in next-to-leading
  order plus parton shower matched simulations of inclusive jet and dijet
  production}}, Phys.Rev. \textbf{D86} (2012),
  \href{http://inspirehep.net/record/1127523}{094042},
  [\href{http://arXiv.org/pdf/1208.2815}{{\tt arXiv:1208.2815}} [hep-ph]].
  \relax
 \relax
\bibitem{Cascioli:2013era}
\href{http://www.slac.stanford.edu/spires/find/hep/www?eprint=1309.5912}{F.~Cascioli,
  P.~Maierhoefer, N.~Moretti, S.~Pozzorini and F.~Siegert}, \emph{{NLO matching
  for $t\bar{t}b\bar{b}$ production with massive $b$-quarks}},
  \href{http://arXiv.org/pdf/1309.5912}{{\tt arXiv:1309.5912}} [hep-ph]. \relax
 \relax
\bibitem{Hoeche:2013mua}
S.~H{\"o}che, J.~Huang, G.~Luisoni, M.~Sch{\"o}nherr and J.~Winter, \emph{{Zero
  and one jet combined NLO analysis of the top quark forward-backward
  asymmetry}}, Phys.Rev. \textbf{D88} (2013),
  \href{http://inspirehep.net/record/1238288}{014040},
  [\href{http://arXiv.org/pdf/1306.2703}{{\tt arXiv:1306.2703}} [hep-ph]].
  \relax
 \relax
\bibitem{Cascioli:2013gfa}
\href{http://www.slac.stanford.edu/spires/find/hep/www?eprint=1309.0500}{F.~Cascioli,
  S.~H{\"o}che, F.~Krauss, P.~Maierhofer, S.~Pozzorini et~al.}, \emph{{Precise
  Higgs-background predictions: merging NLO QCD and squared quark-loop
  corrections to four-lepton + 0,1 jet production}},
  \href{http://arXiv.org/pdf/1309.0500}{{\tt arXiv:1309.0500}} [hep-ph]. \relax
 \relax
\bibitem{Hoche:2010pf}
S.~H{\"o}che, F.~Krauss, M.~Sch{\"o}nherr and F.~Siegert, \emph{{Automating the
  POWHEG method in Sherpa}}, JHEP \textbf{1104} (2011),
  \href{http://www.slac.stanford.edu/spires/find/hep/www?eprint=1008.5399}{024},
   [\href{http://arXiv.org/pdf/1008.5399}{{\tt arXiv:1008.5399}} [hep-ph]].
  \relax
 \relax
\bibitem{Hoeche:2010kg}
S.~H{\"o}che, F.~Krauss, M.~Sch{\"o}nherr and F.~Siegert, \emph{{NLO matrix
  elements and truncated showers}}, JHEP \textbf{08} (2011),
  \href{http://www.slac.stanford.edu/spires/find/hep/www?eprint=arXiv:1009.1127}{123},
   [\href{http://arXiv.org/pdf/1009.1127}{{\tt arXiv:1009.1127}} [hep-ph]].
  \relax
 \relax
\bibitem{Aad:2012en}
G.~Aad et~al., ATLAS Collaboration collaboration, \emph{{Study of jets produced
  in association with a W boson in $pp$ collisions at $\sqrt{s} = 7$ TeV with
  the ATLAS detector}}, Phys.Rev. \textbf{D85} (2012),
  \href{http://inspirehep.net/record/1083318}{092002},
  [\href{http://arXiv.org/pdf/1201.1276}{{\tt arXiv:1201.1276}} [hep-ex]].
  \relax
 \relax
\bibitem{Heister:2003aj}
A.~Heister et~al., ALEPH collaboration, \emph{{Studies of QCD at $e^+ e^-$
  centre-of-mass energies between 91 and 209 GeV}}, Eur. Phys. J. \textbf{C35}
  (2004),
  \href{http://www-spires.dur.ac.uk/spires/find/hep/www?j=EPHJA,C35,457}{457--486}.
  \relax
 \relax
\end{thebibliography}

\end{document}